\documentstyle[12pt,epsfig,times]{article}

\input epsf

\begin{document}

\def\ap{\textrm{'}}
\def\Biops{{\sc Biops} }
\def\Biopsn{{\sc Biops}}
\def\Oops{{\sc Oops} }
\def\Oopsn{{\sc Oops}}
\def\Aixi{{\sc Aixi} }
\def\Aixin{{\sc Aixi}}
\def\tl{{\sc Aixi}{\em (t,l)} }
\def\tln{{\sc Aixi}{\em (t,l)}}
\def\hs{{\sc Hsearch} }
\def\hsn{{\sc Hsearch}}
\def\GM{G\"{o}del Machine }
\def\gm{G\"{o}del machine }
\def\GMn{G\"{o}del Machine}
\def\gmn{G\"{o}del machine}
\newtheorem{theorem}{Theorem}[section]
\newtheorem{lemma}{Lemma}[section]
\newtheorem{corollary}{Corollary}[section]
\newtheorem{conjecture}{Conjecture}[section]
\newtheorem{example}{Example}[section]
\newtheorem{definition}{Definition}[section]
\newtheorem{postulate}{Postulate}[section]
\newtheorem{method}{Method}[section]
\newtheorem{procedure}{Procedure}[section]

\def\odt{{\textstyle{1\over 2}}}
\def\maxarg{\mathop{\rm maxarg}}          
\def\minarg{\mathop{\rm minarg}}          

\begin{titlepage}

\begin{center}

\vspace{3cm}
\vspace{0.6cm}

{\em TR IDSIA-19-03, Version 5 \\
December 2006, arXiv:cs.LO/0309048 v5 \\
(v1: 25 September 2003)
}
\vspace{0.9cm}

{\LARGE \GMn s: Self-Referential \\
Universal Problem Solvers Making \\
\vspace{0.2cm}
Provably Optimal Self-Improvements}

\vspace{0.2cm}

\end{center}

\begin{center}

J\"{u}rgen Schmidhuber

{\it IDSIA, Galleria 2, 6928 Manno-Lugano, Switzerland \& \\
TU M\"{u}nchen, Boltzmannstr. 3,  85748 Garching, M\"{u}nchen, Germany}

{\tt juergen@idsia.ch - http://www.idsia.ch/\~{ }juergen}

\end{center}

\vspace{0.2cm}

\begin{abstract}

We present the first class of mathematically rigorous, general, fully
self-referential, self-improving, optimally efficient problem solvers.
Inspired by Kurt G\"{o}del's celebrated self-referential formulas (1931),
such a problem solver rewrites any part of its own code as soon
as it has found a proof that the rewrite is {\em useful,} where the
problem-dependent utility function and the hardware and the entire
initial code are described by axioms encoded in an initial proof
searcher which is also part of the initial code.  The searcher
systematically and in an asymptotically optimally efficient way tests
computable {\em proof techniques} (programs whose outputs are proofs)
until it finds a provably useful, computable self-rewrite.  We show
that such a self-rewrite is globally optimal---no local maxima!---since
the code first had to prove that it is not useful to continue the
proof search for alternative self-rewrites.  Unlike Hutter's previous {\em
non}-self-referential methods based on hardwired proof searchers, ours
not only boasts an optimal {\em order} of complexity but can optimally
reduce any slowdowns hidden by the $O()$-notation, provided the utility
of such speed-ups is provable at all.

\vspace{0.5cm}

\vspace{0.5cm}
\noindent
{\bf \gm publications} up to 2006
(100 years after Kurt G\"{o}del's birth, and 75 years 
after his landmark paper 
\cite{Goedel:31}
laying the foundations of
theoretical computer science):
 \cite{Schmidhuber:03gm,Schmidhuber:04gmhtml,Schmidhuber:04oops,Schmidhuber:05icann,Schmidhuber:05gmconscious,Schmidhuber:05gmai}.

\end{abstract}

\end{titlepage}

\tableofcontents
\newpage

\section{Introduction and Outline}
\label{intro}

All traditional algorithms for problem solving /
machine learning /
reinforcement learning \cite{Kaelbling:96}
are hardwired. Some are designed to improve some limited
type of policy through
experience, but are not part of the modifiable policy, 
and cannot improve themselves in a theoretically sound way.
Humans are needed to create 
new / better problem solving algorithms and to prove their 
usefulness under appropriate assumptions.

Here we will eliminate the restrictive need for human 
effort in the most general way possible, 
leaving all the work
including the proof search to a system
that can rewrite and improve itself in arbitrary 
computable ways and in a most efficient fashion.
To attack this {\em ``Grand Problem of Artificial Intelligence''}
\cite{Schmidhuber:03grandai},
we introduce a novel class of optimal,
fully self-referential \cite{Goedel:31}
general problem solvers called 
{\em \gmn s} \cite{Schmidhuber:03gm,Schmidhuber:04gmhtml,Schmidhuber:05icann,Schmidhuber:05gmai,Schmidhuber:05gmconscious}.\footnote{Or {\em `Goedel machine'}, to
avoid the {\em Umlaut}. But
{\em `Godel machine'} would not be quite correct.
Not to be confused with what Penrose
calls, in a different context,
{\em `G\"{o}del's putative theorem-proving machine'} \cite{Penrose:94}!}
They are universal problem solving systems that
interact with some (partially observable) environment and 
can in principle modify themselves without essential limits
apart from the limits of computability.  
Their initial algorithm is not hardwired;
it can completely rewrite itself,
but only if a proof searcher embedded within the initial algorithm 
can first prove that the rewrite is useful, 
given a formalized utility function reflecting 
computation time and
expected future success (e.g., rewards).
We will see that self-rewrites due to this approach 
are actually {\em globally optimal} (Theorem \ref{globopt}, 
Section \ref{secglobopt}), relative to G\"{o}del's 
well-known fundamental restrictions of provability \cite{Goedel:31}.
These restrictions should not worry us;
if there is no proof of some self-rewrite's utility, then humans
cannot do much either.

The initial proof searcher is $O()$-optimal 
(has an optimal order of complexity)
in the sense of Theorem \ref{asopt},  Section \ref{biops}. 
Unlike hardwired systems such as Hutter's 
\cite{Hutter:01aixi+,Hutter:01fast+}
and Levin's \cite{Levin:73,Levin:84}
(Section \ref{previous}),
however, a \gm can in principle speed up any part of its
initial software, including its proof searcher, to meet 
{\em arbitrary} formalizable notions of optimality beyond those
expressible in the $O()$-notation.  Our approach yields the first
theoretically sound, fully self-referential,
optimal, general problem solvers.

\noindent {\bf Outline.} Section \ref{basics} presents basic concepts 
and fundamental limitations, Section \ref{details} 
the essential details of a self-referential axiomatic system,
Section \ref{secglobopt} the Global Optimality Theorem \ref{globopt}, 
and Section \ref{biops} the $O()$-optimal 
(Theorem \ref{asopt}) initial proof searcher. 
Section \ref{discussion} provides examples
and relations to previous work,
briefly discusses issues such as 
a {\em technical} justification of consciousness,
and lists answers to several frequently asked questions
about \gmn s.

\section{Overview / Basic Ideas /  Limitations}
\label{basics}

Many traditional problems of computer science require just
one problem-defining input at the beginning of the problem solving
process.  For example, the initial input may be a large integer,
and the goal may be to factorize it.
In what follows, however, we will
also consider the {\em more general case}
where the problem solution requires interaction with a dynamic,
initially unknown environment that produces a continual stream of
inputs and feedback
signals, such as in autonomous robot control tasks,
where the goal may be to maximize expected cumulative
future reward \cite{Kaelbling:96}.
This may require the solution of essentially arbitrary problems
(examples in Section \ref{examples} formulate traditional
problems as special cases).

\subsection{Set-up and Formal Goal}
\label{notation}
Our hardware (e.g., a universal or space-bounded
Turing machine 
\cite{Turing:36} 
or the abstract model
of a personal computer) 
has a single life which
consists of discrete cycles or time steps $t=1, 2, \ldots$.
Its total lifetime $T$ may or may not be known in advance.
In what follows, the value of any time-varying variable $Q$
at time $t$ will be denoted by $Q(t)$.

During each cycle our hardware 
executes an elementary operation which affects its 
variable state $s \in \cal S \subset B^*$
(without loss of generality, $B^*$ is
the set of possible bitstrings over 
the binary alphabet $B=\{ 0, 1 \}$)
and possibly also the variable environmental state $Env \in \cal E$ 
(here we need not yet specify the problem-dependent set $\cal E$).
There is a hardwired
state transition function $F: \cal S \times \cal E \rightarrow \cal S$. 
For $t > 1$, $s(t)=F(s(t-1), Env(t-1))$ is the state at a point where
the hardware operation of cycle $t-1$ is finished, but the one of
$t$ has not started yet.
$Env(t)$ may depend on past output actions encoded in $s(t-1)$ and 
is simultaneously updated or (probabilistically) computed by 
the possibly reactive environment.

In order to talk conveniently about programs and data,
we will often attach names to certain string variables encoded
as components or substrings of $s$.
Of particular interest are the three variables called
{\em time}, {\em x}, {\em y}, and {\em p}:
\begin{enumerate}
\item At time $t$, variable $time$
holds a unique binary representation of $t$.
We initialize $time(1)=`1\ap$, the bitstring consisting only of a one.
The hardware increments $time$ from one cycle to the next.
This requires at most $O(log~t)$ and on average only $O(1)$
computational steps.
 
\item
Variable {\em x} holds the inputs from the environment to the \gmn.
For $t>1$, $x(t)$  may
differ from $x(t-1)$ only if
a program running on the \gm has executed
a special input-requesting instruction at time $t-1$.
Generally speaking, the delays between successive inputs
should be sufficiently large so that
programs can perform
certain elementary computations on an input, such as copying
it into internal storage (a reserved part of $s$)
before the next input arrives.
  
  \item
Variable {\em y} holds the outputs of the \gmn.
$y(t)$ is an output bitstring which may 
subsequently influence the
environment, where $y(1)=`0\ap$ by default.
For example, $y(t)$ could be interpreted as a control
signal for an environment-manipulating
robot whose actions may have an effect on
future inputs.

\item
$p(1)$ is the initial software: a program
implementing the original (sub-optimal) policy 
for interacting with the environment,
represented as a substring $e(1)$ of $p(1)$,
plus the original policy for searching proofs.
Details will be discussed below.
\end{enumerate}

At any given time $t$ ($1 \leq t \leq T$) the goal 
is to maximize future success or {\em utility}.
A typical {\em ``value to go''} utility function
is of the form $u(s, Env): \cal S \times \cal E \rightarrow R$, where
$\cal R$ is the set of real numbers:
\begin{equation}
\label{u}
u(s, Env) =
E_{\mu} \left [ \sum_{\tau=time}^T  r(\tau)~~ \Bigg| ~~s, Env \right ],
\end{equation}
where $r(t)$ is a real-valued reward input (encoded within $s(t)$) at time $t$, 
$E_{\mu}(\cdot \mid \cdot)$ denotes the conditional expectation operator
with respect to some possibly unknown distribution $\mu$ from a set $M$
of possible distributions ($M$ reflects
whatever is known about the possibly probabilistic reactions 
of the environment), and the above-mentioned $time=time(s)$ is a function 
of state $s$ which uniquely identifies the 
current cycle.
Note that we take into account the possibility of extending
the expected lifespan
through appropriate actions.

Alternative formalizable utility functions could favor 
improvement of {\em worst case} instead 
of {\em expected} future performance, 
or higher reward intake {\em per time interval} etc.  
Clearly, most classic problems of computer science
can be formulated in this framework---see examples
in Section \ref{examples}.

\subsection{Basic Idea of \GM}
\begin{figure}[hbt]
\centerline{\epsfig{figure=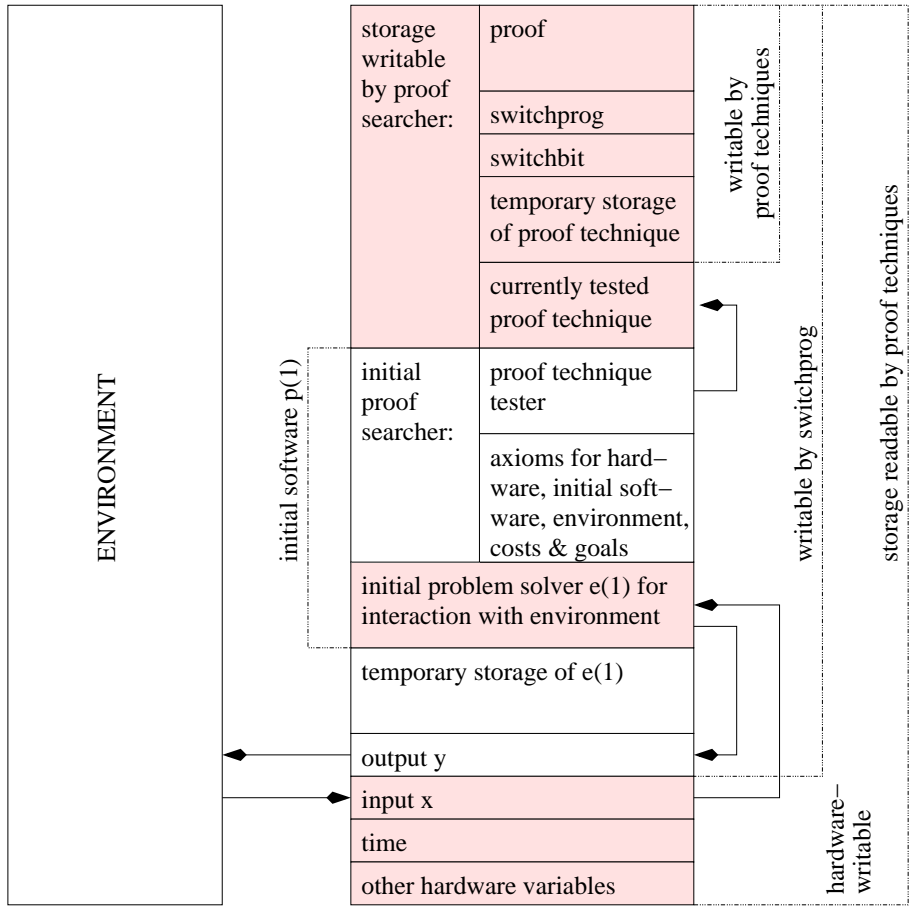,angle=0,height=12cm}}
\caption{{\small
Storage snapshot of a not yet self-improved example \gmn,
with the initial software still intact. See text for details.
}} \label{storage}
\end{figure}

Our machine becomes a self-referential 
\cite{Goedel:31}
{\em \gmn}
by loading it
with a 
particular form of
machine-dependent, 
self-modifying code $p$. The initial code
$p(1)$ at time step 1
includes a (typically sub-optimal)
problem solving subroutine $e(1)$ for interacting with
the environment, such as any traditional reinforcement learning 
algorithm \cite{Kaelbling:96}, 
and a general proof searcher subroutine 
(Section \ref{biops}) 
that systematically makes pairs
{\em (switchprog, proof)} (variable substrings of $s$)
until it finds a {\em proof} of
a target theorem which essentially states: {\em `the
immediate rewrite of {\em p} through current program {\em switchprog}
on the given machine
implies higher utility than leaving {\em p} as is'.} Then it executes
{\em switchprog}, which may completely rewrite $p$, including
the proof searcher. Section \ref{details} will explain
details of the necessary 
initial axiomatic system $\cal A$ 
encoded in $p(1)$.
Compare Fig. \ref{storage}.

\noindent
The {\bf Global Optimality Theorem} (Theorem \ref{globopt}, 
Section \ref{secglobopt}) shows 
this self-im\-prove\-ment strategy is not greedy: since the
utility of {\em `leaving $p$ as is'} implicitly evaluates all possible
alternative {\em switchprog}s which an unmodified $p$ might find later,
we obtain a globally optimal self-change---the {\em current switchprog}
represents the best of all possible relevant self-changes, relative
to the given resource limitations and initial proof search strategy.

\subsection{Proof Techniques and an $O()$-Optimal Initial Proof Searcher}
Section \ref{biops} will present an $O()$-optimal 
initialization of the proof searcher,
that is, one with an optimal {\em order} of complexity
(Theorem \ref{asopt}).  Still, there will remain a lot of 
room for self-improvement hidden by the $O()$-notation. 
The searcher
uses an online extension of {\em Universal Search} 
\cite{Levin:73,Levin:84}
to systematically test {\em online
proof techniques}, which are proof-generating programs that
may read parts of state $s$
(similarly, mathematicians are often more interested in 
proof techniques than in theorems).
To prove target theorems as above,
proof techniques may invoke special instructions
for generating axioms and applying inference rules to prolong the 
current {\em proof} by theorems. Here
an axiomatic system $\cal A$ 
encoded in $p(1)$
includes axioms describing {\bf (a)} how any instruction invoked
by a program running on the given hardware will change 
the machine's state $s$ 
(including instruction pointers etc.)
from one step to the next (such that proof techniques can reason
about the effects of any program including the proof searcher),
{\bf (b)} the initial program $p(1)$ itself (Section \ref{details} will show
that this is possible without introducing circularity),
{\bf (c)} stochastic environmental properties,
{\bf (d)} the formal utility function $u$, 
e.g., equation (\ref{u}), 
which takes into account
computational costs of all actions including proof search.

\subsection{Limitations of \GMn s}
\label{limits}
The fundamental limitations are closely related to
those first identified by G\"{o}del's celebrated paper on
self-referential formulae \cite{Goedel:31}.
Any formal system that encompasses arithmetics (or ZFC etc)
is either flawed or allows for unprovable but true statements.
Hence even a \gm with unlimited computational
resources must ignore those self\--im\-prove\-ments
whose effectiveness it cannot prove,
e.g., for lack of sufficiently powerful axioms in $\cal A$.
In particular, one can construct pathological
examples of environments and
utility functions that make it impossible for the machine
to ever prove a target theorem.
Compare Blum's speed-up theorem 
\cite{Blum:67,Blum:71}
based on certain incomputable predicates.
Similarly, a realistic \gm with limited resources
cannot profit from self\--im\-prove\-ments
whose usefulness it cannot prove within
its time and space constraints.

Nevertheless, unlike previous methods, it can 
in principle exploit at least the {\em provably} good speed-ups 
of {\em any} part of its initial software, including those 
parts responsible for huge (but problem class-independent) slowdowns 
ignored by the earlier approaches \cite{Hutter:01aixi+,Hutter:01fast+}
(Section \ref{previous}).

\section{Essential Details of One Representative \GM}
\label{details}
{\bf Notation.} Unless stated otherwise or obvious, 
throughout the paper newly introduced variables and functions
are assumed to cover the range implicit in the context.
$l(q)$ denotes the number of bits in a bitstring $q$; 
$q_n$ the $n$-th bit of $q$;
$\lambda$ the empty string (where $l(\lambda)=0$);
$q_{m:n}= \lambda$ if $m>n$ and $q_m q_{m+1} \ldots q_n$
otherwise (where $q_0 := q_{0:0} := \lambda$).

Theorem proving requires an 
axiom scheme yielding an 
enumerable set of 
axioms of a formal logic 
system $\cal A$ whose formulas and theorems are symbol 
strings over some finite alphabet that may include traditional
symbols of logic (such as
$\rightarrow,\wedge,=,(,), \forall, \exists, \ldots$,
$c_1, c_2, \ldots,$  $f_1, f_2, \ldots$), 
probability theory (such as $E(\cdot)$, the expectation operator),
arithmetics ($+, -, /, = , \sum, <, \ldots$),
string manipulation (in particular, symbols for 
representing any part of state $s$ at any time, 
such as $s_{7:88}(5555)$).
A proof is a sequence of theorems,
each either an axiom or inferred from previous
theorems by applying one of the inference rules such
as {\em modus ponens} combined with {\em unification}, e.g.,
\cite{Fitting:96}.  

The remainder of this paper will omit standard knowledge to be found 
in any proof theory textbook.
Instead of listing {\em all} axioms of a particular $\cal A$ in
a tedious fashion,
we will focus on the novel and critical details: 
how to overcome potential problems with self-reference 
and how to deal with the potentially delicate online generation of proofs 
that talk about and affect the currently running proof generator itself. 

\subsection{Proof Techniques}
\label{prooftech}
Brute force proof searchers
(used in Hutter's work \cite{Hutter:01aixi+,Hutter:01fast+}; 
see Section \ref{previous} for a review)
systematically generate all proofs
in order of their sizes.  To produce a certain proof,
this takes time exponential in proof size.
Instead our $O()$-optimal $p(1)$ will produce many proofs
with low algorithmic complexity 
\cite{Solomonoff:64,Kolmogorov:65,LiVitanyi:97}
much more quickly. It systematically tests (see Section \ref{biops})
programs called
{\em proof techniques} written in universal language
$\cal L$ implemented within $p(1)$.  
For example, $\cal L$  may be a variant of PROLOG \cite{Prolog:87}
or the universal {\sc Forth}\cite{Forth:70}-inspired
programming language used in recent work on optimal search
\cite{Schmidhuber:04oops}.
A proof technique is composed
of instructions that allow any part of $s$ to be read, 
such as inputs encoded in variable $x$ (a substring of $s$) or
the code of $p(1)$. It may write on $s^p$, a  part of $s$ reserved for 
temporary results.  It also may rewrite {\em switchprog}, 
and produce an incrementally growing proof placed in the
string variable {\em proof} stored somewhere in $s$.
{\em proof} and $s^p$ are reset to the empty string at 
the beginning of each new proof technique test.
Apart from standard arithmetic and function-defining instructions 
\cite{Schmidhuber:04oops} that modify $s^p$,
the programming language $\cal L$ includes special instructions
(details in Section \ref{instructions})
for prolonging the current {\em proof} by correct theorems,
for setting {\em switchprog},
and for checking whether a provably optimal $p$-modifying
program was found and should be executed now.
Certain long proofs can be produced by short proof techniques.

\subsection{Important Instructions Used by Proof Techniques}
\label{instructions}

The nature of the six {\em proof}-modifying instructions below
(there are no others)  
makes it impossible to insert an 
incorrect theorem into {\em proof},
thus trivializing proof verification.
Let us first provide a brief overview of the most
important instructions: {\bf get-axiom(n)}
appends the $n$-th possible axiom
to the current {\em proof}, 
{\bf apply-rule(k, m, n)} applies the $k$-th
inference rule to the $m$-th and $n$-th theorem
in the current {\em proof} (appending the result),
{\bf set-switchprog(m,n)}
sets $switchprog:= s^p_{m:n}$, 
and {\bf check()}
tests whether the last theorem in {\em proof}
is a {\bf target theorem} showing that a self-rewrite
through {\em switchprog} would be useful.
The details are as follows.

\begin{enumerate}

\item {\bf get-axiom(n)}
takes as argument an integer $n$ computed 
by a prefix of the currently tested proof technique
with the help of arithmetic instructions 
such as those used in previous work
\cite{Schmidhuber:04oops}. 
Then it appends
the $n$-th axiom (if it exists, according to the axiom scheme below) 
as a theorem to the current theorem
sequence in {\em proof}.  The initial axiom scheme encodes:

\begin{enumerate}
\item
\label{hardwareaxioms}
{\bf Hardware axioms}
describing the hardware,
formally specifying how 
certain components of $s$ (other than
environmental inputs $x$) may
change from one cycle to the next. 

For example, if the hardware is a
Turing machine\footnote{
Turing reformulated G\"{o}del's unprovability results in terms
of Turing machines (TMs) \cite{Turing:36} which
subsequently became the most widely used abstract model of
computation.  It is well-known that there are {\em universal}
TMs that in a certain sense can emulate any other TM or any
other known computer.  G\"{o}del's integer-based formal language
can be used to describe any universal TM, and vice versa.}
(TM) \cite{Turing:36},
then $s(t)$ is a bitstring that encodes the current
contents of all tapes of the TM, the positions of its
scanning heads, and the current {\em internal state}
of the TM's finite state automaton, while $F$ specifies
the TM's look-up table which maps any possible combination
of internal state and bits above scanning heads
to a new internal state and an action such as:
replace some head's current bit by 1/0, increment
(right shift) or decrement (left shift) some
scanning head, read and copy next input bit to cell above
input tape's scanning head,
etc.  

Alternatively, if the hardware is given by the abstract model
of a modern microprocessor with limited storage, $s(t)$ will
encode the current storage contents,
register values,
instruction pointers etc.

For instance, the following axiom could
describe how some 64-bit hardware's instruction pointer
stored in $s_{1:64}$ is continually incremented as long as 
there is no overflow and the value
of $s_{65}$ does not indicate that a jump to some other address
should take place:
\[
(\forall t \forall n :
[(n < 2^{64}-1) \wedge  (n > 0) \wedge  (t > 1) \wedge  (t < T) 
\]
\[
\wedge 
(string2num(s_{1:64}(t))=n) 
\wedge (s_{65}(t)=`0\ap) ]
\]
\[
\rightarrow
(string2num(s_{1:64}(t+1))=n+1))
\]
Here the semantics of used symbols such 
as `('  and `$>$'  and `$\rightarrow$' (implies)
are the traditional ones, while `$string2num$'
symbolizes a function
translating bitstrings into numbers.
It is clear that any abstract hardware model can be
fully axiomatized in a similar way.

\item
{\bf Reward axioms}
defining the computational costs of any
hardware instruction, 
and physical costs of output 
actions, such as control signals $y(t)$
encoded in $s(t)$.
Related axioms assign values to certain input events 
(encoded in variable $x$, a substring of $s$)
representing reward or punishment (e.g.,
when a \gmn-controlled robot bumps into an obstacle).
Additional axioms define the total value of the \gmn 's life as a 
scalar-valued function of all 
rewards (e.g., their sum) and 
costs experienced between cycles $1$ and $T$, etc.
For example, assume that
$s_{17:18}$ can be changed only through
external inputs; the following example axiom 
says that the total reward increases by 3 whenever
such an input equals `11'
(unexplained symbols carry the obvious meaning):
\[
(\forall t_1 \forall t_2:
[(t_1 < t_2) \wedge  (t_1 \geq 1) 
\wedge  (t_2 \leq T)  
\wedge  (s_{17:18}(t_2)=`11\ap)  ]
\]
\[
\rightarrow
[R(t_1,t_2)= R(t_1,t_2-1)+3]),
\]
where $R(t_1,t_2)$ is interpreted as 
the cumulative reward between times $t_1$ and $t_2$.
It is clear that any formal scheme for producing
rewards can be fully axiomatized in a similar way.

\item
\label{envaxioms}
{\bf Environment axioms}
restricting the way the environment will produce 
new inputs (encoded within certain substrings of $s$) in reaction to 
sequences of outputs $y$ encoded in $s$.
For example, it may be known
in advance that the environment is sampled from an unknown probability
distribution $\mu$ contained in a given set $M$
of possible distributions (compare equation \ref{u}).
E.g., $M$ may contain all distributions
that are computable, given the previous history
\cite{Solomonoff:64,Solomonoff:78,Hutter:01aixi+},
or at least limit-computable \cite{Schmidhuber:00v2,Schmidhuber:02ijfcs}.
Or, more restrictively, the environment
may be some unknown but deterministic computer
program \cite{Zuse:69,Schmidhuber:97brauer}
sampled from the Speed Prior \cite{Schmidhuber:02colt} which assigns
low probability to environments that are hard to compute by any method.
Or the interface to the environment is Markovian \cite{Schmidhuber:91nips},
that is, the current
input always uniquely identifies the environmental state---a lot
of work has already been done 
on this special case \cite{Samuel:59,Bellman:61,Sutton:98}.
Even more restrictively, the environment may evolve in completely
predictable fashion known in advance.
All such prior assumptions
are perfectly formalizable in an appropriate $\cal A$ 
(otherwise we could not write scientific papers about them).

\item
{\bf Uncertainty axioms; string manipulation axioms:}
\label{probaxioms}
Standard axioms for arithmetics and calculus 
and probability theory \cite{Kolmogorov:33} 
and statistics 
and string manipulation that (in conjunction
with the hardware axioms and environment axioms) allow for constructing proofs 
concerning (possibly uncertain) properties of future values of 
$s(t)$ as well as bounds on expected remaining lifetime /
costs / rewards,  
given some time $\tau$ and certain 
hypothetical values for components of $s(\tau)$ etc. 
An example theorem saying something about
expected properties of future inputs $x$ might look like this:
\[
(\forall t_1 \forall \mu \in M:
[(1 \leq t_1) \wedge
(t_1 + 15597 < T) \wedge
(s_{5:9}(t_1) = `01011\ap)
\]
\[
\wedge (x_{40:44}(t_1)=`00000\ap) ]
\rightarrow
(\exists t: [(t_1 < t < t_1 + 15597)  
\]
\[
\wedge
(P_{\mu}(x_{17:22}(t) = `011011\ap \mid s(t_1)) > \frac{998}{1000} )])),
\]
where $P_{\mu}(. \mid . )$ represents a conditional probability
with respect to an axiomatized prior distribution $\mu$ from
a set of distributions $M$ described by
the environment axioms (Item \ref{envaxioms}).

Given a particular 
formalizable hardware 
(Item \ref{hardwareaxioms})
and formalizable assumptions about the 
possibly probabilistic environment
(Item \ref{envaxioms}),
obviously one can fully axiomatize 
everything that is needed for 
proof-based reasoning about past and future machine states.

\item
\label{initaxioms}
{\bf Initial state axioms:}
Information about how
to reconstruct the initial state $s(1)$ or parts thereof,
such that the proof searcher 
can build proofs including
axioms of the type
\[
(s_{{\bf m}:{\bf n}}(1)={\bf z}),
~e.g.:~ (s_{7:9}(1)=`010\ap).
\]
Here and in the remainder of the paper 
we use bold font in formulas to indicate 
syntactic place holders (such as {\bf m,n,z})
for symbol strings representing 
variables (such as {\em m,n,z})
whose semantics are explained in 
the text---in the present context $z$ is the 
bitstring $s_{m:n}(1)$.

Note that it is {\bf no fundamental problem} to fully encode
both the hardware description {\em and} the initial
hardware-describing $p$ within $p$
itself. To see this, observe that some software may 
include a program that can print the software.

\item
{\bf Utility axioms}
\label{itemu}
describing the overall goal
in the form of utility function $u$; e.g., equation (\ref{u}) in
Section \ref{notation}.
\end{enumerate}

\item
\label{applyrule} 
{\bf apply-rule(k, m, n)}
takes as arguments the index $k$ (if it exists) of 
an inference rule 
such as {\em modus ponens} 
(stored in a list of possible inference rules
encoded within $p(1)$) and the indices $m, n$ 
of two previously proven theorems (numbered in order of
their creation) in the current {\em proof}.
If applicable, the corresponding inference rule is 
applied to the addressed theorems
and the resulting theorem appended to {\em proof}. Otherwise
the currently tested proof technique is interrupted.  This 
ensures that {\em proof} is never fed with invalid proofs. 

\item
{\bf delete-theorem(m)}
deletes the $m$-th theorem in the currently stored
{\em proof}, thus freeing storage such that proof-storing parts of $s$
can be reused and the maximal proof size is not necessarily limited by
storage constraints.
Theorems deleted from {\em proof}, however,  cannot be addressed any more by
{\em apply-rule} to produce further prolongations of {\em proof}.

\item
{\bf set-switchprog(m,n)}
replaces $switchprog$ by $s^p_{m:n}$, provided
that $s^p_{m:n}$ is indeed a non-empty substring of
$s^p$, the storage writable by proof techniques.

\item 
\label{check}
{\bf check()}
verifies whether the goal of the proof search has been reached.
First it tests whether the last theorem (if any) in {\em proof}
has the form of a {\bf target theorem}. A target theorem states
that given the {\em current} axiomatized utility
function $u$ (Item \ref{itemu}), the utility of a
switch from $p$ to the current {\em switchprog}
would be higher than the utility of continuing the execution
of $p$ (which would keep searching for alternative {\em switchprog}s).
Target theorems are
symbol strings 
(encoded in $s$) 
of the (decoded) form
\begin{equation}
\label{goal}
(u[s({\bf t_1}) \oplus (switchbit({\bf t_1})=`1\ap), Env({\bf t_1})] >
\]
\[
u[s({\bf t_1}) \oplus (switchbit({\bf t_1})=`0\ap), Env({\bf t_1})] )
\end{equation}
where the variable $t_1$ (represented by syntactic place holder ${\bf t_1}$)
stands for a time step,
while all other symbols belong to the alphabet of
the theorem-proving calculus, that is,
the set of possible target theorems is parameterized only by $t_1$.
Here the calculus should permit the notation
$s({\bf t_1}) \oplus (switchbit({\bf t_1})=`b\ap)$ as
a shortcut for the state obtained when we replace
{\em switchbit}($t_1$), 
the true value of the variable bit $switchbit$
(encoded in $s$) 
at time $t_1$, by $b \in \{0, 1\}$. This will
facilitate the formulation of theorems
that compare values conditioned on various alternative
hypothetical properties of $s(t_1)$. (Note that $s(t_1)$ may
be only partially known by the current proof technique
even in environments where
$s(t_1)$ and {\em switchbit}($t_1$) are
fully predetermined for all valid $t_1$.)

The purpose of introducing $t_1$ is to deal with hardware-specific
temporal delays that may be involved in checking and switching---it may
take a significant amount of
time to match abstract symbol strings found during proof search to the
\gmn's real current state.
If a target theorem has been found, {\em check()} uses a
simple prewired subroutine (also encoded in $p(1)$, of course)
to check whether there is enough time left
to set variable {\em switchbit} (originally 0) to 1 before
the continually increasing $time$ will equal $t_1$.
If this subroutine
returns a negative result, {\em check()} exits.
Otherwise it sets {\em switchbit} $:=1$
(there is no other way of changing {\em switchbit}).
Then it repeatedly tests $time$ until $time > t_1$, to make sure
the condition of formula (\ref{goal})
was fulfilled at $t_1$.
Then it transfers control to {\em switchprog}
(there is no other way of calling {\em switchprog}).
The {\em switchprog}
may subsequently rewrite all parts of $s$, excluding hardware-reserved
parts such as $time$ and $x$, but including $p$.

\item 
\label{state2theorem} 
{\bf state2theorem(m, n)}
takes two integer arguments $m, n$
and tries to transform the current contents of $s_{m:n}$ 
into a theorem of the form 
\[
(s_{{\bf m}:{\bf n}}({\bf t_1})={\bf z}),
~e.g.:~ (s_{6:9}(7775555)=`1001\ap),
\]
where $t_1$ represents a time measured (by checking {\em time})
shortly after {\em state2theorem} was invoked, 
and $z$ the bistring $s_{m:n}(t_1)$ (recall the special 
case $t_1=1$ of Item \ref{initaxioms}).
So we accept the time-labeled current 
observable contents of any part of $s$ as a theorem that does not have
to be proven in an alternative way from, say, the 
initial state $s(1)$, because the computation so far
has already demonstrated that the theorem is true.
Thus we may exploit information conveyed
by environmental inputs, and the fact that sometimes
(but not always) the fastest way
to determine the output of a program is to run it.

{\em
This non-traditional online interface between syntax and semantics
requires special care though.
We must avoid inconsistent results through
parts of $s$ that change while being read.
For example, the present value of a quickly changing
instruction pointer {\em IP} (continually updated by the hardware)
may be essentially unreadable in the
sense that the execution of the reading subroutine
itself will already modify {\em IP} many times.
For convenience, the (typically limited) hardware could be set up
such that it stores the contents of
fast hardware variables every $c$ cycles in
a reserved part of $s$, such that
an appropriate  variant of {\em state2theorem()} could at least
translate certain recent values of fast variables into theorems.
This, however, will not abolish {\em all} problems associated
with self-observations.
For example, the $s_{m:n}$ to be read might
also contain the reading procedure's
own, temporary, constantly changing string pointer variables,
etc.\footnote{We see that certain parts of the
current $s$ may not be directly observable without changing
the observable itself.
Sometimes, however, axioms and previous observations will allow
the \gm to {\em deduce} time-dependent storage contents that
are not directly observable.
For instance, by analyzing the code being executed through
instruction
pointer {\em IP} in the example above,  the value of {\em IP} at
certain times may be predictable (or postdictable, after the
fact).  The values of other variables at given times,
however, may not be deducible at all.
Such limits of self-observability
are reminiscent of Heisenberg's celebrated
uncertainty principle \cite{Heisenberg:25},
which states that certain physical measurements are necessarily
imprecise, since the measuring process affects the measured
quantity.}
To address such problems on computers with limited
memory, {\em state2theorem}
first uses some fixed protocol
(encoded in $p(1)$, of course)
to check whether the current $s_{m:n}$ is readable
at all or whether it might change if it
were read by the remaining code of {\em state2theorem}.
If so, or if $m,n,$ are not in the proper range,
then the instruction has no further effect. Otherwise it
appends an {\em observed} theorem of the form
$(s_{{\bf m}:{\bf n}}({\bf t_1})={\bf z})$
to {\em proof}.
For example, if the current time is 7770000, then the invocation
of {\em state2theorem(6,9)} might return the theorem $(s_{6:9}(7775555)=`1001\ap)$,
where
$7775555-7770000=5555$ reflects the time needed by {\em state2theorem}
to perform the initial check
and to read leading bits off the continually increasing $time$
(reading $time$ also costs time) such that
it can be sure that $7775555$ is a recent proper time label following
the start of {\em state2theorem}.
}

\end{enumerate}

The axiomatic system $\cal A$ is a defining
parameter of a given \gmn.  Clearly,  $\cal A$ must be
strong enough to permit proofs of target theorems.
In particular, the theory of uncertainty axioms
(Item \ref{probaxioms}) must be sufficiently rich.
This is no fundamental problem: we simply
insert all traditional axioms of probability theory \cite{Kolmogorov:33}.

\section{\bf Global Optimality Theorem}
\label{secglobopt}
 
 Intuitively, at any given time $p$ should
 execute some self-modification algorithm 
 (via instruction {\em check()}---Item \ref{check} above) 
 only if it is
 the `best' of all possible self-modifications,
 given the utility function, which typically
 depends on available resources, such as storage size and
 remaining lifetime.
 At first glance, however, target theorem (\ref{goal})
 seems to implicitly talk about just
 one single modification algorithm, namely, {\em switchprog}$(t_1)$
 as set by the systematic proof searcher at time $t_1$.
 Isn't this type of local search greedy? Couldn't
 it lead to a local optimum instead of a global one?
 No, it cannot, according to the following global optimality theorem.

\subsection{Globally Optimal Self-Changes, Given $u$ and $\cal A$ Encoded in $p$}
\label{sectheorem}
  
  \begin{theorem}
  \label{globopt}
  Given any formalizable utility function $u$ (Item \ref{itemu}),
  and assuming consistency of the underlying formal system $\cal A$,
  any self-change of $p$ obtained through execution of
  some program {\em switchprog} identified
  through the proof of a target theorem  (\ref{goal})
  is globally optimal in the following sense:
  the utility of starting the execution of the present
  {\em switchprog} is higher than the utility of
  waiting for the proof searcher
  to produce an alternative {\em switchprog} later.
  \end{theorem}

\noindent
{\bf Proof.} Target theorem  (\ref{goal})
implicitly talks about all the other
{\em switchprog}s that the proof searcher
could produce in the future. To see this, consider
the two alternatives of the binary decision:
(1) either execute the current {\em switchprog} (set {\em switchbit} $=1$), or
(2) keep searching for {\em proof}s and {\em switchprog}s (set {\em switchbit} $=0$)
until the systematic
searcher comes up with an even better {\em switchprog}.
Obviously the second alternative concerns all (possibly infinitely
many) potential {\em switchprog}s to be considered later.  That is,
if  the current {\em switchprog}  were not the `best', then
the proof searcher would not be able to prove that
setting {\em switchbit} and
executing  {\em switchprog} will cause higher expected reward
than discarding {\em switchprog}, assuming consistency of $\cal A$.
{\em Q.E.D.}

The initial proof searcher of Section \ref{biops}
already generates all possible proofs and {\em switchprogs} in $O()$-optimal fashion. 
Nevertheless, since it is part of $p(1)$, its proofs can speak about the
proof searcher itself, possibly triggering proof searcher rewrites 
resulting in better than merely $O()$-optimal performance.

\subsection{Alternative Relaxed Target Theorem}
\label{alternative}
We may replace the target theorem  (\ref{goal}) (Item \ref{check})
by the following 
alternative target theorem:
\[
(u[s({\bf t_1}) \oplus (switchbit({\bf t_1})=`1\ap), Env({\bf t_1})] \geq
\]
\begin{equation}
\label{goal2}
u[s({\bf t_1}) \oplus (switchbit({\bf t_1})=`0\ap), Env({\bf t_1})] )
\end{equation}
The only difference to the original target theorem  (\ref{goal})
is that the ``$>$'' sign became a  ``$\geq$'' sign. That is,
the \gm will change itself as soon as it has found a proof
that the change will not make things worse.
A Global Optimality Theorem similar to Theorem \ref{globopt}
holds; simply replace the last phrase 
in Theorem \ref{globopt} by:
  {\em the utility of starting the execution of the present
  {\em switchprog} is at least as high as the 
  utility of waiting for the proof searcher
  to produce an alternative {\em switchprog} later.}

\subsection{Global Optimality and Recursive Meta-Levels}
\label{comment}
One of the most important
aspects of our fully self-referential set-up is the following.
Any proof of a target theorem automatically proves 
that the corresponding self-modification is good for all
further self-modifications affected by the present one,
in recursive fashion. 
In that sense all possible ``meta-levels'' of the self-referential
system are collapsed into one. 

\subsection{How Difficult is it to Prove Target Theorems?}
\label{difficult}
This depends on the tasks and the initial axioms $\cal A$, of course.
It is straight-forward to devise simple tasks and 
corresponding consistent $\cal A$
such that there are short and trivial proofs of target theorems.

Even when we initialize the initial problem solver $e(1)$ by an
asymptotically optimal, rather general method such as Hutter's \tl 
\cite{Hutter:01aixi+,Hutter:04book+},
it may be straight-forward to prove that switching to
another strategy is useful,
especially when $\cal A$ contains additional prior
knowledge in form of axiomatic assumptions 
beyond those made by \tln .  The latter needs a very time-consuming but constant set-up
phase whose costs disappear in the $O()$-notation but not in 
a utility function such as the $u$ of equation (\ref{u}). 
For example, simply construct an environment where
maximal reward is achieved by performing a never-ending sequence
of simple but rewarding actions, say, repeatedly pressing a lever,
plus a very simple axiomatic system $\cal A$ that permits a short proof
showing that it is useful to interrupt the non-rewarding set-up phase and
start pressing the lever. 

On the other hand, it is possible to construct
situations where it is impossible to prove target theorems,
for example, by using results of undecidability theory, 
e.g., \cite{Goedel:31,Rice:53,Blum:67,Blum:71}.
In particular, adopting the extreme notion of triviality
embodied by Rice's theorem \cite{Rice:53}
({\em any nontrivial property over general
functions is undecidable}), only {\em trivial}
improvements of a given strategy may be provably useful.
This notion of triviality, however, clearly does not reflect
what is {\em intuitively} regarded as trivial by scientists.
Although many theorems of the machine learning
literature in particular,  and the computer science 
literature in general, are limited to functional properties that
are trivial in the sense of Rice, they are widely regarded 
as non-trivial in an {\em intuitive} sense. 
In fact, the infinite domains of function classes 
addressed by Rice's theorem 
are irrelevant not only for most scientists dealing
with real world problems 
but also for a typical \gm dealing
with a limited number of events that
may occur within its limited life time. 
Generally speaking, in between the {\em obviously} trivial and the {\em obviously}
non-trivial cases there are many less obvious ones.
The point is: usually we do not know in advance whether it is
possible or not to change a given initial problem solver
in a provably good way.  The traditional approach is to 
invest human research effort into finding out. A \gmn, however, can 
do this by itself, without essential limits
apart from those of computability and provability.  

Note that to prove a target theorem, 
a proof technique does not necessarily have to compute
the true expected utilities of switching and not
switching---it just needs to determine which is higher.
For example, it may be easy to prove that
speeding up a subroutine of the proof searcher
by a factor of 2 will certainly be worth
the negligible (compared to lifetime $T$) time needed to
execute the subroutine-changing algorithm, no matter
what is the precise utility of the switch.

\section{Bias-Optimal Proof Search (BIOPS)}
\label{biops}

Here we construct an initial  $p(1)$ that is $O()$-optimal in a certain
limited sense to be described below, but still might be improved
as it is not necessarily optimal in the sense of the given $u$
(for example, the $u$ of equation (\ref{u}) neither mentions
nor cares for $O()$-optimality).
Our Bias-Optimal Proof Search (BIOPS)
is essentially an application of 
{\em Universal Search}
\cite{Levin:73,Levin:84}
to proof search. 
One novelty, however, is this:
Previous practical variants and extensions of {\em Universal
Search} have been applied
\cite{Schmidhuber:95kol,Schmidhuber:97nn,Schmidhuber:97bias,Schmidhuber:04oops}
to {\em offline}
program search tasks where the program inputs are fixed
such that the same program always produces the same results.
In our {\em online} setting, however, BIOPS has to take
into account that the same proof technique 
started at different times may yield different proofs,
as it may read parts of $s$ (e.g., inputs)
that change as the machine's life proceeds. 

\subsection{Online Universal Search in Proof Space}
\label{online}

BIOPS starts with a probability distribution $P$ 
(the initial bias) on the proof techniques $w$ that
one can write in $\cal L$,
e.g., $P(w)=K^{-l(w)}$ for programs composed from $K$
possible instructions  \cite{Levin:84}.
BIOPS is {\em near-bias-optimal} 
\cite{Schmidhuber:04oops}
in the sense that it will not spend 
much more time on any proof technique than it deserves,
according to its probabilistic bias,
namely, not much more than its probability times the total search time:
\begin{definition} [Bias-Optimal
Searchers \cite{Schmidhuber:04oops}]
\label{bias-optimal}
{\em Let $\cal R$ be a problem class,
$\cal C$ be a search space of solution candidates
(where  any problem $r \in \cal R$ should have a solution in $\cal C$),
$P(q \mid r)$ be a task-dependent bias in the form of conditional probability
distributions on the candidates $q \in \cal C$. Suppose that we also have
a predefined procedure that creates and tests any given $q$
on any $r \in \cal R$ within time $t(q,r)$ (typically unknown in advance).
Then} a searcher is $n$-bias-optimal ($n \geq 1$) if
for any maximal total search time $T_{total} > 0$
it is guaranteed to solve any problem $r \in \cal R$
if it has a solution $p \in \cal C$
satisfying $t(p,r) \leq P(p \mid r)~T_{total}/n$.
It is bias-optimal if  $n=1$.
\end{definition}

\begin{method}[BIOPS]
\label{lsearch}
{\em 
In phase $(i=1,2,3, \ldots)$ {\sc Do}:
{\sc For} all self-delimiting \cite{Levin:84} 
proof techniques $w \in \cal L$ satisfying $P(w) \geq 2^{-i}$ {\sc Do}:
\begin{enumerate}
\item
Run $w$ until halt or error (such as 
division by zero) or $2^iP(w)$ steps consumed. 
\item
Undo effects of $w$ on $s^p$
(does not cost significantly more time 
than executing $w$).
\end{enumerate}
}
\end{method}
A proof technique $w$ can interrupt Method \ref{lsearch} 
only by invoking instruction {\em check()} (Item \ref{check}), 
which may transfer
control to {\em switchprog} (which possibly
even will delete or rewrite Method  \ref{lsearch}).
Since the initial $p$ 
runs on the formalized hardware, and since proof techniques
tested by $p$ can read $p$ and other parts of $s$, they
can produce proofs concerning the (expected) 
performance of $p$ and BIOPS itself.
Method \ref{lsearch} at least has the 
optimal {\em order} of computational complexity in the following
sense. 
\begin{theorem}
\label{asopt}
If independently of variable {\em time(s)} some unknown 
fast proof technique $w$
would require at most $f(k)$ steps to produce
a proof of difficulty measure $k$ (an integer depending on
the nature of the task to be solved), then
Method \ref{lsearch}
will need at most $O(f(k))$ steps.
\end{theorem}
{\bf Proof.}
It is easy to see that
Method \ref{lsearch} will need at most $O(f(k)/P(w)) = O(f(k))$
steps---the constant factor  $1/P(w)$ does not depend on $k$.
{\em Q.E.D.}

The initial proof search itself is merely $O()$-optimal.
Note again, however,  that
the proofs themselves may concern quite
different, arbitrary formalizable notions of optimality
(stronger than those expressible in the $O()$-notation)
embodied by the given, problem-specific, formalized 
utility function $u$, in particular, the maximum future
reward in the sense of equation (\ref{u}).
This may provoke useful, constant-affecting rewrites of 
the initial proof searcher despite its limited (yet popular
and widely used) notion of $O()$-optimality.
Once a useful rewrite has been found and executed
after some initial fraction of the \gmn 's total lifetime, the restrictions
of $O()$-optimality need not be an issue any more.

\subsection{How a Surviving Proof Searcher May
Use the Optimal Ordered Problem Solver to Solve Remaining Proof Search Tasks}
\label{remaining}

The following is not essential for this paper.
Let us assume that the execution of the {\em switchprog}
corresponding to the first found target theorem has not rewritten the code of
$p$ itself---the current $p$ is still equal to $p(1)$---and has
reset {\em switchbit} and returned
control to $p$ such that it can continue where it was interrupted.
In that case the \Biops subroutine of $p(1)$ can use the Optimal
Ordered Problem Solver \Oops
\cite{Schmidhuber:04oops}
to accelerate the search for the
$n$-th target theorem ($n>1$) by reusing proof techniques for earlier found
target theorems where possible.  The basic ideas are as
follows (details: \cite{Schmidhuber:04oops}).

Whenever a target theorem has been proven, $p(1)$ {\em freezes}
the corresponding proof technique: it becomes non-writable
by proof techniques to be tested in later proof search tasks,
but remains readable,
such that it can be copy-edited and/or invoked as a subprogram by future proof
techniques.  We also allow prefixes of proof
techniques to temporarily rewrite the probability distribution
on their suffixes
\cite{Schmidhuber:04oops},
thus essentially rewriting
the probability-based search
procedure (an incremental extension of Method \ref{lsearch})
based on previous experience. As a side-effect we metasearch
for faster search procedures, which can greatly accelerate the learning of
new tasks \cite{Schmidhuber:04oops}.

Given a new proof search task, \Biops performs \Oops
by spending half the total search time on a
variant of
Method \ref{lsearch}
that searches only among self-delimiting
\cite{Levin:74,Chaitin:75} proof techniques starting with the most recently
frozen proof technique.  The rest of the
time is spent on fresh proof techniques with arbitrary prefixes
(which may reuse previously frozen proof techniques though)
\cite{Schmidhuber:04oops}.
(We could also search for a {\em generalizing} proof technique
solving all proof search tasks so far. In the first half of the search
we would not have to test proof techniques on tasks other than
the most recent one, since we already know that their prefixes
solve the previous tasks
\cite{Schmidhuber:04oops}.)

It can be shown that \Oops is
essentially {\em 8-bias-optimal} (see Def. \ref{bias-optimal}),
given either the initial bias or intermediate biases due to
frozen solutions to previous tasks
\cite{Schmidhuber:04oops}.
This result immediately carries over to \Biopsn.
To summarize, \Biops essentially allocates part of the total search
time for a new task to proof techniques that exploit previous successful
proof techniques in computable ways.  If the new task can be solved faster
by copy-editing / invoking previously frozen
proof techniques than by solving the new proof search task from scratch,
then  \Biops will discover this and profit thereof. If not, then
at least it will not be significantly slowed down by the previous
solutions---\Biops will remain 8-bias-optimal.
 
 Recall, however, that \Biops is not the only
 possible way of initializing the \gmn's proof searcher.
The Global Optimality Theorem \ref{globopt} 
(Section \ref{secglobopt})
expresses optimality with respect to whichever
initial proof searcher we choose.

\section{Discussion \& Previous Work}
\label{discussion}

Here we list a few examples of
possible types of self-improvements
(Section \ref{possible}),
\gm applicability to various tasks defined by various utility functions and
environments
(Section \ref{examples}),
probabilistic hardware 
(Section \ref{prob}),
and relations to previous work
(Section \ref{previous}).
We also briefly discuss self-reference 
and consciousness 
(Section \ref{conscious}),
and provide a list
of answers to frequently asked questions 
(Section \ref{faq}).

\subsection{Possible Types of \GM Self-Improvements}
\label{possible}

  Which provably useful self-modifications are possible?
  There are few limits to what a \gm  might do.
   
\begin{enumerate}
\item
   In one of the simplest cases
   it might leave its basic proof searcher intact
   and just change the ratio of time-sharing between
   the proof searching subroutine and the
   subpolicy $e$---those parts of $p$
   responsible for interaction with the environment.
    
\item
    Or the \gm might modify $e$ only.
    For example, the initial $e(1)$ may be a program
    that regularly stores limited memories
    of past events somewhere in $s$; this might allow $p$ to derive that
    it would be useful to modify $e$ such that $e$ will conduct certain
    experiments to increase the knowledge about
    the environment, and use the resulting information
    to increase reward intake.  In this sense the \gm embodies
    a principled way of dealing with the
    exploration vs exploitation problem \cite{Kaelbling:96}.
    Note that the {\em expected} utility (equation (\ref{u}))
    of conducting some experiment may  exceed
    the one of not conducting it,
    even when the experimental outcome later suggests to
    keep acting in line with the previous $e$.

\item
The \gm might also modify its very axioms
to speed things up. For example,
it might find a proof that the
original axioms should be replaced or
augmented by theorems derivable
from the original axioms.
 
\item
 The \gm might even change
 its own utility function and target theorem,
 but can do so only if their {\em new} values
 are provably better according to the {\em old} ones.
  
\item
  In many cases we do not expect
  the \gm to replace its proof searcher by
  code that completely abandons the search for proofs.
  Instead we expect that only certain subroutines
  of the proof searcher will be sped up---compare the
  example in Section \ref{difficult}---or that perhaps just
  the order of generated proofs will be
  modified in problem-specific fashion. This could be done
  by modifying
  the probability distribution on the proof techniques of
  the initial bias-optimal proof searcher
  from Section \ref{biops}.

\item
\label{supermarket}
Generally speaking,
the utility of limited rewrites may often be
easier to prove than the one of total rewrites.
For example, suppose it is 8.00pm and our \gmn-controlled
agent's permanent goal is to
maximize future expected reward, using the (alternative)
target theorem (\ref{goal2}).  Part thereof is to avoid hunger. There
is nothing in its fridge, and shops close down at 8.30pm.  It does not
have time to optimize its way to the supermarket in  every little detail,
but if it does not get going right now it will stay hungry tonight (in
principle such near-future consequences of actions should be easily
provable, possibly even in a way related to
how humans prove advantages of potential actions to themselves). That
is, if the agent's previous policy did not already include, say, an automatic
daily evening trip to the supermarket, the policy provably should be
rewritten at least in a very limited and simple way right now, while there
is still time, such that the agent will surely get some food tonight,
without affecting less urgent future behavior that  can be optimized
/ decided later, such as details of the route to the food, or
of tomorrow's actions.
 
\item
 In certain uninteresting environments reward
 is maximized by becoming dumb. For example,
 a given task may require to repeatedly
 and forever execute the same pleasure center-activating action,
 as quickly as possible.  In such cases the \gm may delete
 most of its more time-consuming initial software
 including the proof searcher.
  
\item
  Note that there is no reason why a \gm should not
  augment its own hardware.
  Suppose its lifetime is known to be 100 years.
  Given a hard problem and
  axioms restricting the possible
  behaviors of the environment,
  the \gm might find a proof that its
  expected cumulative reward will increase if
  it invests 10 years into building faster computational
  hardware, by exploiting the physical resources of
  its environment.
\end{enumerate}

\subsection{Example Applications}
\label{examples}

Traditional examples that
do not involve significant interaction with a probabilistic
environment are easily dealt with in our reward-based framework:

\begin{example}[Time-limited NP-hard optimization]
\label{tsp}
The initial
input to the \gm is the representation of
a connected graph with
a large number of nodes linked by
edges of various lengths.
Within given time $T$ it should
find a cyclic path connecting all nodes.
The only real-valued reward will occur
at time $T$. It equals
1 divided by the length
of the best path found so far (0 if none was found).
There are no other inputs.
The by-product of  maximizing
expected reward is to find the shortest
path findable within the limited time,
given the initial bias.
\end{example}
 
 \begin{example}[Fast theorem proving]
 \label{goldbach}
 Prove or disprove as quickly as possible
 that all even integers  $>2$ are the sum of
 two primes (Goldbach's conjecture).
 The reward is $1/t$,
 where $t$ is the time required to produce
 and verify the first such proof.
 \end{example}
More general cases are:

\begin{example}[Maximizing expected reward with bounded resources]
\label{robot}
A robot that needs at least
1 liter of gasoline per hour
interacts with a partially unknown environment,
trying to find hidden, limited gasoline depots
to occasionally refuel its tank.
It is rewarded in proportion to its lifetime,
and dies after at most 100 years or as
soon as its tank is empty or
it falls off a cliff etc.
The probabilistic environmental reactions are initially
unknown but assumed to be sampled from
the axiomatized Speed Prior \cite{Schmidhuber:02colt},
according to which
hard-to-compute environmental reactions are unlikely.
This permits a computable strategy for making near-optimal
predictions  \cite{Schmidhuber:02colt}.
One by-product of maximizing  expected reward
is to maximize expected lifetime.
\end{example}
\begin{example}[Optimize any suboptimal problem solver]
\label{optimize}
Given any formalizable problem, implement
a suboptimal but known problem solver
as software on the \gm hardware,
and let the proof searcher of Section \ref{biops} run in parallel.
\end{example}

\subsection{Probabilistic \GM Hardware}
\label{prob}
Above we have focused on an example deterministic machine
living in a possibly probabilistic environment.
It is straight-forward to extend this
to computers whose actions are computed in
probabilistic fashion, given the current state.
Then the expectation calculus
used for probabilistic aspects of the environment
simply has to be extended to the hardware itself,
and the mechanism for verifying proofs has to
take into account that there is no such thing as
a certain theorem---at best there are formal statements
which are true with such and such probability.
In fact, this may be the most realistic approach
as any physical hardware is error-prone,
which should be taken into account by
realistic probabilistic \gmn s.
 
 Probabilistic settings also automatically avoid certain
 issues of axiomatic consistency. For example, predictions
 proven to come true with
 probability less than 1.0 do not necessarily cause contradictions
 even when they do not match the observations.

\subsection{Relations to Previous Work}
\label{previous}
 
Despite (or maybe because of) the ambitiousness and
potential power of self-improving machines,
there has been little work in this vein outside our own
labs at IDSIA and TU M\"{u}nchen.
Here we will list essential differences
between the \gm and our previous approaches to `learning to learn,'
`metalearning,' self-improvement, self-optimization,  etc.

The most closely related approaches are Hutter's
\hs and \tl (Item \ref{hutter} below).
For historical reasons, however, we will first discuss
Levin's {\em Universal Search} and  Hutter's \Aixin.
\begin{enumerate}

\item{\bf \GM vs Universal Search}
\label{gmvsls}

Unlike the fully self-referential \gmn, 
Levin's {\em Universal Search} \cite{Levin:73,Levin:84}
has a hardwired, unmodifiable meta-algorithm that cannot improve itself. It
is asymptotically optimal for inversion problems whose solutions can be quickly
verified in $O(n)$ time (where $n$ is the solution size), but it will always
suffer from the same huge constant slowdown factors (typically $>> 10^{1000}$)
buried in the $O()$-notation. The self-improvements of a \gmn,
however, can be more than merely $O()$-optimal, since its utility function
may formalize a stonger type of optimality that does not ignore huge
constants just because they are constant---compare the utility
function of equation (\ref{u}).

Furthermore, the \gm is applicable to 
general lifelong reinforcement learning (RL) tasks \cite{Kaelbling:96}
where {\em Universal Search} is {\em not} asymptotically optimal,
and not even applicable, since
in RL the evaluation of 
some behavior's value in principle consumes the learner's entire life!
So the naive test of whether a program is good or not
would consume the entire life. That is, we could test
only one program; afterwards life would be over.

Therefore, to achieve their objective, general RL machines 
must do things that {\em Universal Search} does
not do, such as predicting {\em future} tasks and rewards.
This partly motivates Hutter's universal RL machine AIXI, to be
discussed next.
\\

\item{\bf \GM vs \Aixi}
\label{aixi}

Unlike \gmn s,
Hutter's recent {\em \Aixi model} \cite{Hutter:01aixi+,Hutter:04book+}
generally needs {\em unlimited} computational resources per input update.
It combines Solomonoff's universal prediction
scheme \cite{Solomonoff:64,Solomonoff:78} with an {\em expectimax}
computation.  In discrete cycle $k=1,2,3, \ldots$,
action $y(k)$ results
in perception $x(k)$ and reward $r(k)$, both
sampled from the unknown
(reactive) environmental probability distribution $\mu$.  \Aixi defines
a mixture distribution $\xi$ as a weighted sum of distributions $\nu
\in \cal M$, where $\cal M$ is any class of distributions that includes the
true environment $\mu$.  For example, $\cal M$ may
be a sum of all computable
distributions \cite{Solomonoff:64,Solomonoff:78}, where the sum of
the weights does not exceed 1.  In cycle $k+1$, \Aixi
selects as next action the first in an action sequence maximizing
$\xi$-predicted reward up to some given horizon.
Recent work \cite{Hutter:02selfopt+} demonstrated \Aixi's optimal
use of observations as follows.  The Bayes-optimal policy $p^\xi$ based on
the mixture $\xi$ is self-optimizing in the sense that its average
utility value converges asymptotically for all $\mu \in \cal M$ to the
optimal value achieved by the (infeasible) Bayes-optimal policy $p^\mu$
which knows $\mu$ in advance.  The necessary condition that $\cal M$
admits self-optimizing policies is also sufficient.
Furthermore, $p^\xi$ is Pareto-optimal
in the sense that there is no other policy yielding higher or equal
value in {\em all} environments $\nu \in \cal M$ and a strictly higher
value in at least one \cite{Hutter:02selfopt+}.

While \Aixi clarifies certain
theoretical limits of machine learning, it
is computationally intractable, especially when $\cal M$
includes all computable
distributions.  This drawback
motivated work on the time-bounded, asymptotically optimal
\tl system \cite{Hutter:01aixi+}
and the related \hs  \cite{Hutter:01fast+}, both
to be discussed next.
\\

\item{\bf \GM vs \hs and \tl}
\label{hutter}

Now we come to the most closely related previous work;
so we will go an extra length to 
point out the main novelties of the \gmn.

Hutter's non-self-referential but still $O()$-optimal 
{\em `fastest' algorithm for all well-defined problems}
\hs \cite{Hutter:01fast+} 
uses a {\em hardwired} brute force proof searcher  and
ignores the costs of proof search.
Assume discrete input/output domains $X/Y$, a formal problem
specification $f: X \rightarrow Y$
(say, a functional description of how integers are decomposed
into their prime factors),
and a particular $x \in X$ (say,
an integer to be factorized). \hs
orders all proofs of an appropriate axiomatic system
by size to find programs $q$ that
for all $z \in X$ provably compute $f(z)$
within time bound $t_q(z)$. Simultaneously it
spends most of its time on executing the $q$ with the
best currently proven time bound $t_q(x)$.
It turns out that \hs
is as fast as the {\em fastest} algorithm that provably
computes $f(z)$ for all $z \in X$, save for a constant factor
smaller than $1 + \epsilon$ (arbitrary $\epsilon > 0$)
and an $f$-specific but $x$-independent
additive constant  \cite{Hutter:01fast+}.
This constant may be enormous though.

Hutter's \tl \cite{Hutter:01aixi+}
is related.
In discrete cycle $k=1,2,3, \ldots$ of \tln's lifetime,
action $y(k)$ results in perception $x(k)$ and reward $r(k)$,
where all quantities may depend on the complete history.
Using a universal computer such as a Turing machine,
\tl needs an initial offline
setup phase (prior to interaction with the environment) where
it uses a {\em hardwired} brute force proof searcher to
examine all proofs of length at
most $L$, filtering out those that identify programs (of maximal
size $l$ and maximal runtime $t$ per cycle) which not only
could interact with the environment but which for
all possible interaction histories
also correctly predict a lower bound of their own expected future reward.
In cycle $k$, \tl then runs all programs
identified in the
setup phase (at most $2^l$), finds the one with highest self-rating,
and executes its corresponding action.
The problem-independent setup time (where almost all of the work is done)
is $O(L \cdot 2^{L})$.  The online time per cycle is $O(t \cdot 2^l)$.
Both are constant but typically huge.

\noindent
{\bf Advantages and Novelty of the \GMn.}
There are major differences between the \gm and Hutter's
\hs \cite{Hutter:01fast+} and \tl \cite{Hutter:01aixi+}, including:

\begin{enumerate}
\item 
The theorem provers  of \hs and \tl 
are hardwired, non-self-\-refe\-ren\-tial, unmodifiable meta-algorithms 
that cannot improve
themselves. That is, they will always suffer from the same 
huge constant slowdowns (typically $\gg 10^{1000}$)
buried in the $O()$-notation.
But there is nothing in principle that prevents the
truly self-referential code
of a \gm
from proving and exploiting drastic
reductions of such constants, in the best possible 
way that provably constitutes an improvement, if there is any. 
\item
The demonstration of the $O()$-optimality of \hs and \tl depends on 
a clever allocation of computation time to some of their unmodifiable meta-algorithms.
Our Global Optimality Theorem (Theorem \ref{globopt}, 
Section \ref{secglobopt}), however, is justified through
a quite different type of reasoning which indeed exploits and crucially depends
on the fact that there is no unmodifiable software at all,
and that the proof searcher itself is readable, modifiable,
and can be improved. 
This is also the reason why its self-improvements
can be more than merely $O()$-optimal.
\item
\hs uses a ``trick'' of proving more than is necessary
which also disappears in the sometimes quite misleading $O()$-notation: it
wastes time on finding programs that provably compute $f(z)$
for all $z \in X$ even when the current $f(x) (x \in X)$ is the only
object of interest. A \gmn, however, needs to prove only what is 
relevant to its goal formalized by $u$. For example, the 
general $u$ of eq. (\ref{u}) 
completely ignores the limited concept of $O()$-optimality, 
but instead formalizes a stronger type of optimality that does not ignore huge
constants just because they are constant.
\item
Both the \gm and \tl can maximize expected reward
(\hs cannot). But the \gm is more flexible
as we may plug in {\em any} type of formalizable utility function
(e.g., {\em worst case} reward),  and unlike \tl it does not
require an enumerable environmental distribution. 
\end{enumerate}
Nevertheless, we may 
use \tl or \hs or other less general methods
to initialize the
substring $e$ of $p$ which is
responsible for interaction with the environment.
The \gm will replace $e(1)$ as soon
as it finds a provably better strategy.

It is the {\em self-referential} aspects
of the \gm that relieve us of much of the burden of careful
algorithm design required for \tl and \hsn. They make the \gm both
conceptually simpler {\em and} more general.
\\

\item{\bf \GM vs \Oops}
\label{oopsrl}
 
The Optimal  Ordered Problem Solver
\Oops \cite{Schmidhuber:04oops,Schmidhuber:03nips}
(used by \Biops in Section \ref{remaining}) extends
{\em Universal Search} (Item \ref{gmvsls}). It is a bias-optimal (see Def. \ref{bias-optimal})
way of searching for a
program that solves each problem in an ordered  sequence
of problems of a rather general type,
continually organizing and managing and reusing earlier acquired knowledge.
Solomonoff recently also proposed related
ideas for a {\em scientist's assistant}
\cite{Solomonoff:03} that modifies the probability
distribution of {\em Universal Search} \cite{Levin:73} based
on experience.

Like {\em Universal Search} (Item \ref{gmvsls}), 
\Oops is not directly applicable to
RL problems.  A provably optimal RL machine
must somehow {\em prove} properties of otherwise un-testable behaviors
(such as: what is the expected reward of this behavior which
one cannot naively test as there is not enough time).
That is part of what the \gm does: it tries
to greatly cut testing time, replacing naive time-consuming
tests by much faster proofs of predictable test outcomes
whenever this is possible.

Proof verification
itself can be performed very quickly. In particular,
verifying the correctness of
a found proof typically does not consume the remaining life.
Hence the \gm may use \Oops as a bias-optimal proof-searching
submodule (Section \ref{remaining}). Since the proofs themselves may concern quite
different, {\em arbitrary} notions of optimality (not just
bias-optimality), the \gm is more
general than plain \Oopsn.
But it is not just an extension of \Oopsn.  Instead of \Oops it may
as well use non-bias-optimal alternative methods to initialize
its proof searcher.
On the other hand, \Oops is not just a precursor of the \gmn.
It is a stand-alone, incremental, bias-optimal way of allocating
runtime to programs that reuse previously successful
programs, and is applicable to many traditional
problems, including but not limited to proof search.
\\

\item{\bf \GM vs Success-Story Algorithm and Other Metalearners}
\label{ssa}

A learner's modifiable components are called its policy. An algorithm
that modifies the policy is a learning algorithm. If the learning
algorithm has modifiable components represented as part of the policy,
then we speak of a self-modifying policy (SMP) \cite{Schmidhuber:97ssa}.
SMPs can modify the way they modify themselves etc.
The \gm has an SMP.

In previous practical work we used the {\em success-story algorithm} (SSA)
to force some (stochastic) SMP to trigger better and better
self-modifications
\cite{Schmidhuber:94self,Schmidhuber:96meta,Schmidhuber:97ssa,Schmidhuber:97bias}.
During the learner's life-time, SSA is occasionally called at times computed
according to SMP itself. SSA uses backtracking to undo those SMP-generated
SMP-modifications that have not been empirically observed to trigger
lifelong reward accelerations (measured up until the current SSA
call---this evaluates the long-term effects of SMP-modifications setting the
stage for later SMP-modifications). SMP-mo\-di\-fi\-ca\-tions that survive SSA
represent a lifelong success history. Until the next SSA call, they build
the basis for additional SMP-modifications. Solely by self-modifications
our SMP/SSA-based learners solved a complex task in a partially observable
environment whose state space is far bigger than most found in
the literature \cite{Schmidhuber:97ssa}.
 
 The \gmn's training algorithm is theoretically
 much more powerful than SSA though.
 SSA empirically measures the usefulness of previous
 self-modifications, and does not necessarily encourage
 provably optimal ones.
 Similar drawbacks hold for
 Lenat's human-assisted, non-autonomous,
 self-modifying learner \cite{Lenat:83},
 our Meta-Genetic Programming \cite{Schmidhuber:87} extending
 Cramer's Genetic Programming \cite{Cramer:85,Banzhaf:98},
 our metalearning economies \cite{Schmidhuber:87}
 extending Holland's machine learning eco\-no\-mies \cite{Holland:85},
 and gradient-based metalearners
 for continuous program spaces of differentiable
 recurrent neural networks
 \cite{Schmidhuber:93selfreficann,Hochreiter:01meta}.
 All these methods, however, could be used to seed
 $p(1)$ with an initial policy.
 \\
\end{enumerate}

\subsection{Are Humans Probabilistic \GMn s?}
\label{human}
We do not know. We think they better be.
Their initial underlying formal system for dealing with uncertainty
seems to differ substantially from those of
traditional expectation calculus and logic though---compare Items
\ref{envaxioms} and \ref{probaxioms} in Section \ref{instructions} as
well as the supermarket example 
(Item \ref{supermarket} in Section \ref{possible}).

\subsection{\GMn s and Consciousness}
\label{conscious}

In recent years the topic of consciousness
has gained some credibility as a serious research issue,
at least in philosophy and neuroscience, e.g., \cite{Crick:98}.
However, there is a lack 
of {\em technical} justifications of consciousness:
so far nobody has shown that 
consciousness is really useful for solving problems,
although problem solving is considered 
of central importance in philosophy \cite{Popper:99}. 

The fully self-referential \gm may be viewed as providing 
just such a technical justification \cite{Schmidhuber:05gmconscious}.
It is ``conscious'' or ``self-aware'' in the sense that
its entire behavior is open to self-introspection,
and modifiable.  
It may `step outside of itself' \cite{Hofstadter:79} 
by  executing self-changes that are provably good,
where the proof searcher itself is subject
to analysis and change through the proof techniques it tests.
And this type of total self-reference is precisely 
the reason for its optimality as a problem solver, in the 
sense of Theorem \ref{globopt}.

\subsection{Frequently Asked Questions}
\label{faq}

In the past year the author frequently
fielded questions about the \gmn. 
Here a list of answers to typical ones.

\begin{enumerate}

\item
{\bf Q:} {\em 
Does the exact business of formal proof search 
really make sense in the uncertain real world?
}

{\bf A:} 
Yes, it does. We just need to insert into $p(1)$ the standard axioms 
for representing uncertainty and for dealing with
probabilistic settings and expected rewards etc. 
Compare items 
\ref{probaxioms} and \ref{envaxioms}
in Section \ref{instructions}, and the definition
of utility as an {\em expected} value in equation (\ref{u}).
Also note that the machine learning literature is
full of {\em human}-generated proofs of properties 
of methods for dealing with stochastic environments.

\item
{\bf Q:} {\em 
The target theorem (\ref{goal}) seems to refer only
to the {\em very first} self-change, which  may completely 
rewrite the proof-search subroutine---doesn't this
make the proof of Theorem \ref{globopt} invalid?
What prevents later self-changes from being destructive?
}

{\bf A:} 
This is fully taken care of.
Please have a look once more at the 
proof of Theorem \ref{globopt},
and note that the first self-change will
be executed only if it is provably useful 
(in the sense of the present untility function $u$)
for all
future self-changes (for which the present self-change is setting
the stage). This is actually one of the main points of
the whole self-referential set-up.

\item
{\bf Q} (related to the previous item): {\em 
The \gm implements a meta-learning behavior: what about a 
meta-meta, and a meta-meta-meta level? 
}

{\bf A:} 
The beautiful thing is that all meta-levels are 
automatically collapsed into one: any proof
of a target theorem automatically proves 
that the corresponding self-modification is good for all
further self-modifications affected by the present one,
in recursive fashion. Recall Section \ref{comment}.

\item
{\bf Q:} {\em 
The \gm software can produce only computable mappings
from input sequences to output sequences.
What if the environment is non-computable?
}

{\bf A:} 
Many physicists and other scientists
(exceptions: \cite{Zuse:69,Schmidhuber:97brauer})
actually seem to assume the real world
makes use of all the real numbers, most
of which are incomputable.
Nevertheless,
theorems and proofs are just finite symbol strings,
and all treatises of physics contain only computable 
axioms and theorems, even when some of the theorems
can be interpreted as making statements about uncountably 
many objects, such as all the real numbers. (Note though that
the L\"{o}wenheim-Skolem Theorem \cite{Loewenheim:15,Skolem:19}
implies that any first order theory with an uncountable model
such as the real numbers also has a countable model.) 
Generally speaking, formal descriptions of non-computable objects
do {\em not at all} present a fundamental problem---they 
may still allow for finding a strategy
that provably maximizes utility. If so, a \gm
can exploit this. If not, then humans will not
have a fundamental advantage over \gmn s.

\item
{\bf Q:} {\em 
Isn't automated theorem-proving 
very hard?  Current AI systems cannot prove 
nontrivial theorems without human intervention 
at crucial decision points.
}

{\bf A:} 
More and more important mathematical proofs (four color 
theorem etc) heavily depend on automated proof search.
And traditional theorem provers do not even make use of
our novel notions of proof techniques and $O()$-optimal
proof search.  Of course, some proofs are indeed hard to find, 
but here humans and \gmn s face the 
same fundamental limitations.

\item
{\bf Q:} {\em Don't the
``no free lunch theorems'' \cite{Wolpert:97}
say that it is 
impossible to construct universal problem solvers?
}

{\bf A:} 
No, they do not. They refer to the very special 
case of problems sampled from {\em i.i.d.} uniform 
distributions on {\em finite} problem spaces. 
See the discussion of no free lunch theorems
in an earlier paper \cite{Schmidhuber:04oops}.

\item
{\bf Q:} {\em 
Can't the \gm
switch to a program {\em switchprog} that rewrites 
the utility function to a ``bogus'' utility function that 
makes unfounded promises of big rewards in the near future?
}

{\bf A:} 
No, it cannot. It should be obvious that rewrites
of the utility function can happen only if
the \gm first can prove that the rewrite
is useful according to the {\em present} utility function.

\item
{\bf Q:} {\em 
Aren't there problems with undecidability? For example,
doesn't Rice's theorem \cite{Rice:53} 
or Blum's speed-up theorem \cite{Blum:67,Blum:71}
pose problems?
}

{\bf A:} Not at all.  Of course, the \gm cannot profit
from a hypothetical useful self-improvement whose utility is undecidable,
and will therefore simply ignore it. Compare Section 
\ref{limits} on fundamental
limitations of \gmn s (and humans, for that matter).
Nevertheless, unlike previous methods, a \gm can in principle
exploit at least the provably good improvements and speed-ups of 
{\em any} part of its initial software.

\end{enumerate}

\section{Conclusion}
\label{conclusion}

In 1931, Kurt G\"{o}del laid the foundations of theoretical
computer science, using
elementary arithmetics to build a universal programming language
for encoding arbitrary proofs, given an arbitrary enumerable set of
axioms.  He went on to construct {\em
self-referential} formal statements that claim their own unprovability,
using Cantor's diagonalization trick \cite{Cantor:1874}
to demonstrate that formal systems such as
traditional mathematics are either flawed in a certain
sense or contain unprovable but true statements \cite{Goedel:31}.
Since G\"{o}del's exhibition of the fundamental limits of proof and
computation, and Konrad Zuse's subsequent construction of the first working
programmable computer (1935-1941),
there has been a lot of work on specialized algorithms
solving problems taken from more or less general problem classes.
Apparently, however,
one remarkable fact has so far escaped the attention of computer
scientists: it is possible to use self-referential proof systems to
build optimally efficient yet conceptually very simple
universal problem solvers.

The initial software $p(1)$ of our \gm
runs an initial, typically sub-optimal problem solver, e.g., one of
Hutter's approaches 
\cite{Hutter:01aixi+,Hutter:01fast+}
which have at least an optimal {\em order} of complexity,
or some less general method  \cite{Kaelbling:96}.
Simultaneously, it runs an $O()$-optimal initial proof 
searcher using an online variant of {\em Universal Search}
to test {\em proof techniques}, which are 
programs able to compute proofs
concerning the system's own future performance,
based on an axiomatic system $\cal A$ 
encoded in $p(1)$,
describing a formal {\em utility} function $u$,
the hardware and $p(1)$ itself. 
If there is no provably good, globally optimal 
way of rewriting $p(1)$ at all, then humans
will not find one either. But if there is one,
then $p(1)$ itself can find 
and exploit it. This approach yields the first
class of theoretically sound, fully self-referential,
optimally efficient, general problem solvers.

After the theoretical discussion in Sections \ref{intro}-\ref{biops},
one practical question remains: to build a particular,
especially practical \gm with small initial constant overhead,
which generally useful 
theorems should one add as axioms to $\cal A$
(as initial bias) such that the initial searcher  
does not have to prove them from scratch? 

\section{Acknowledgments}
Thanks to
Alexey Chernov,
Marcus Hutter,
Jan Poland,
Ray Solomonoff,
Sepp Hochreiter,
Shane Legg,
Leonid Levin,
Alex Graves,
Matteo Gagliolo,
Viktor Zhumatiy,
Ben Goertzel,
Will Pearson,
and Faustino Gomez,
for useful comments on
drafts or summaries or
earlier versions of this paper.
I am also grateful to many others who 
asked questions during \gm talks or
sent comments by email. Their input helped to shape 
Section \ref{faq} on frequently asked questions.

\newpage
\bibliography{bib}
\bibliographystyle{plain}
\end{document}